\documentclass[article,onecolumn,11pt]{IEEEtran}
\usepackage{tikz}
\usepackage{pgfplots}
\usepackage{mathtools}
\mathtoolsset{showonlyrefs=true}
\usepackage{scrextend}
\usetikzlibrary{dsp,chains}

\usepackage{subcaption}
\usepackage{framed}
\usepackage{enumerate}
\usepackage[utf8]{inputenc} 
\usepackage[T1]{fontenc}
\usepackage{url}
\usepackage{cite}
\usepackage{amsmath,amssymb}
\usepackage{epstopdf}
\usepackage{epsfig}
\usepackage{calc,pstricks, pgf, xcolor}
\usepackage{bbding}
\usepackage{bbm}
\usepackage{dsfont}
\usepackage{centernot}
\usepackage{bm}
\usepackage{float}

\newcommand{\dB}{\mathrm{dB}}

\newcommand{\snr}{\mathrm{SNR}}
\newcommand{\bsnr}{\mathrm{S}\wt{\mathrm{N}}\mathrm{R}}

\newcommand{\Pe}{p_e}

\newcommand{\dfn}{\triangleq}
\def\wt{\widetilde}

\begin{document}
\title{Simple Modulo can {Significantly} Outperform Deep Learning-based Deepcode}

\author{Assaf~Ben-Yishai and Ofer~Shayevitz
\thanks{
Assaf Ben-Yishai is with the School of Computer Science and Engineering, Hebrew University of Jerusalem, Israel. 
Ofer Shayevitz is with the Department of EE--Systems, Tel-Aviv University, Tel-Aviv, Israel. Emails: \{assafbster@gmail.com, ofersha@eng.tau.ac.il\}.}}

\date{}
\maketitle
\begin{abstract}
\textit{Deepcode} (H.~Kim et~al.~2018 \cite{kim2018deepcode}) is a recently suggested Deep Learning-based scheme for communication over the AWGN channel with noisy feedback, claimed to be superior to all previous schemes in the literature. Deepcode's use of nonlinear coding (via Deep Learning) has been inspired by known shortcomings (Y.-H.~Kim~et~al 2007 \cite{Kim2007}) of linear feedback schemes. In  2014, we presented a nonlinear feedback coding scheme based on a combination of the classical Schalwijk-Kailath (SK) scheme and modulo-arithmetic \cite{SimpleInteractionAllerton2014}, using a small number of elementary operations without any type of neural network. This \textit{Modulo-SK} scheme has been omitted from the performance comparisons made in the Deepcode paper \cite{kim2018deepcode}, due to its use of common randomness (dither), and in a later version~\cite{DeepcodeIEEE} since it was incorrectly interpreted as a variable-length coding scheme. However, the dither in Modulo-SK was used only for the standard purpose of tractable performance analysis, and is not required in practice. In this short note, we show that a fully-deterministic Modulo-SK (without any dithering) can outperform Deepcode. For example, to attain an error probability of $10^{-4}$ at rate $1/3$ and feedforward $\snr$ of $0\dB$, Modulo-SK requires $3\dB$ less feedback $\snr$ than Deepcode. To attain an error probability of $10^{-6}$ in the same setup but with noiseless feedback, Deepcode requires $150$ rounds of communication, whereas Modulo-SK requires only $15$ rounds, even if the feedback is noisy (with $27\dB$ $\snr$). 

We further address the numerical stability issues of the original SK scheme reported in the Deepcode paper, and explain how they can be avoided. We augment this report with an online-available, fully-functional Matlab simulation for both the classical and Modulo-SK schemes \cite{ModuloSKcode}. Finally, note that Modulo-SK is by no means claimed to be the best possible solution; in particular, using deep learning in conjunction with modulo-arithmetic might lead to better designs, and remains a fascinating direction for future research. 
\end{abstract}
\nocite{DeepcodeNIPS}
\section{AWGN with Noisy Feedback}
Terminal A and Terminal B are connected by a pair of independent AWGN channels, given by:
\begin{align}
Y_n=X_n+Z_n ,\quad \wt{Y}_n=\wt{X}_n+\wt{Z}_n,
\end{align}
where $X_n, Y_n$ (resp. $\wt{X}_n,\wt{Y}_n$) are the input and output of the feedforward (resp. feedback) channel at time $n$, respectively. The feedforward (resp. feedback) channel noise sequence $Z_n\sim \mathcal{N}(0,\sigma^2)$ (resp. $\wt{Z}_n\sim \mathcal{N}(0,\wt{\sigma}^2)$) is i.i.d., and independent of the input $X_n$ (resp. $\wt{X}_n$). Terminal A wants to send a message $W$ to Terminal B over a fixed number of communication rounds $N$, where $W$ is uniformly distributed over a set of cardinality $M$. To that end, at time $n$, the terminals send: 
\begin{align}
  X_n=\varphi_n(W,\wt{Y}^{n-1}), \quad \wt{X}_n=\wt{\varphi}_n(Y^n), 
\end{align}
where $\varphi,\wt{\varphi}$ are predetermined functions, 
such that the average power constraint $P$ (resp. $\wt{P}$) are satisfied: 
\begin{align}
\sum_{n=1}^N\mathbb{E}(X_n^2) \leq N P, \quad \sum_{n=1}^N\mathbb{E}(\wt{X}_n^2) \leq N \wt{P} .
\end{align}
This formulation is referred to in the literature as \textit{active} feedback, whereas \textit{passive} feedback is the special case where $\wt{X}_n=Y_n$. Modulo-SK uses active feedback, while to the best of our understanding, Deepcode uses passive feedback. We denote the feedforward (resp. feedback) signal-to-noise ratio by $\snr\dfn\frac{P}{\sigma^2}$ 
(resp.  $\bsnr\dfn \frac{\wt{P}}{\wt{\sigma}^2}$). 
A communication scheme $(\varphi,\wt{\varphi})$ achieves a rate $R\dfn \frac{\log{M}}{N}$ and an error probability $\Pe$, which is the probability that Terminal B errs in decoding the message $W$ at time $N$, under the optimal decision rule.
{In the sequel we denote the number of information bits by $K\triangleq \log M=N\cdot R$.}
\newline
\section{Results}
\begin{figure*}[!ht]
    \centering
    \begin{subfigure}[t]{0.5\textwidth}
        \centering
        \includegraphics[height=2.7in]{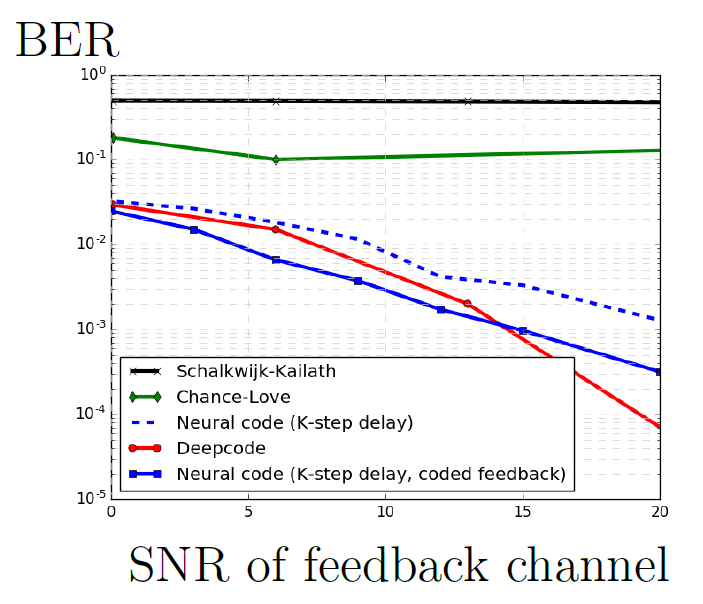}
        \caption{
        \textbf{Deepcode. \cite[Figure~5 (Left)]{kim2018deepcode}}
        \label{fig:Figure5LeftDeepcode}
        }
    \end{subfigure}%
    ~ 
    \begin{subfigure}[t]{0.5\textwidth}
        \centering
        \includegraphics[height=2.7in]{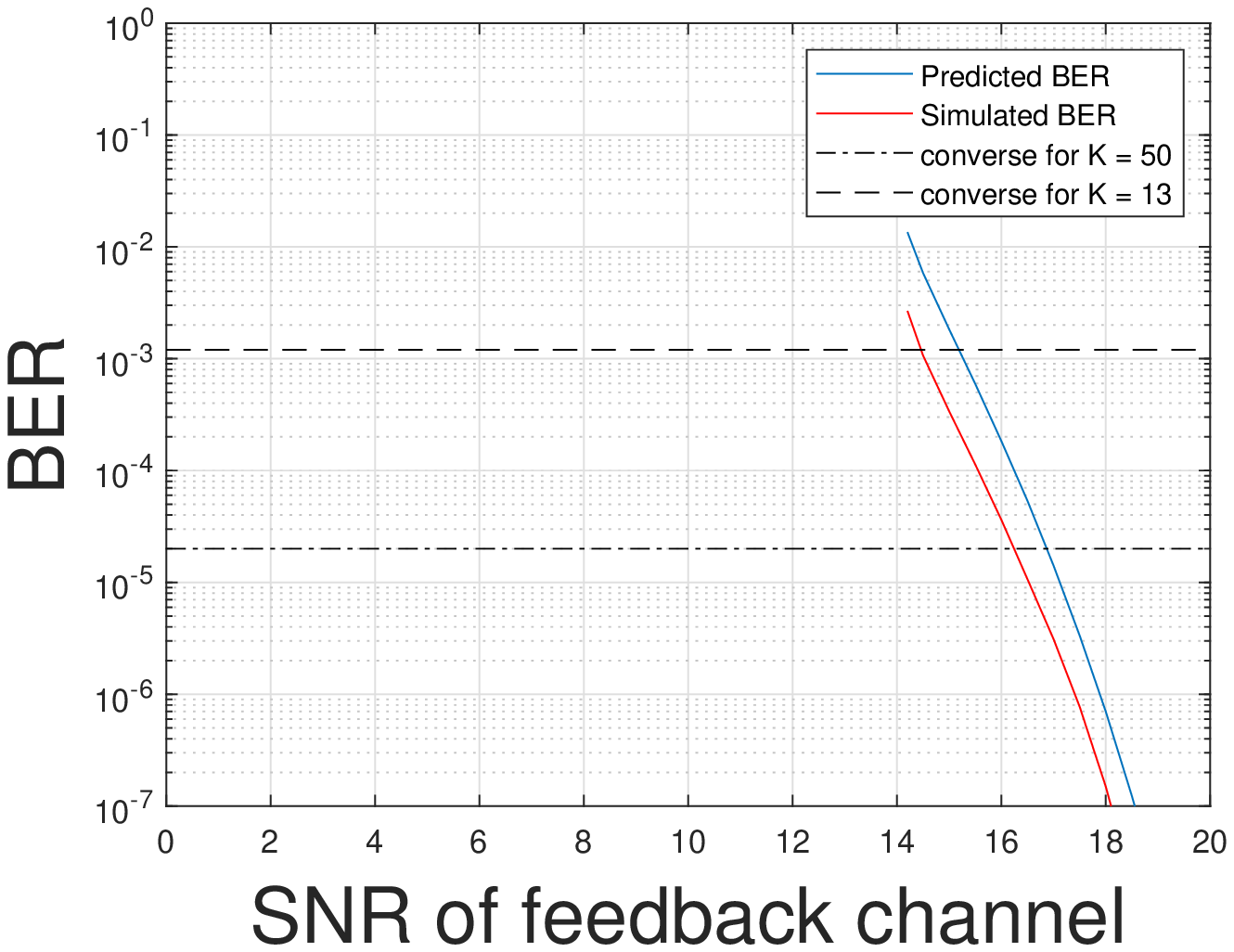}
        \caption{\textbf{Modulo-SK}
        \label{fig:Figure5LeftMSK}
        }
     \end{subfigure}
    \caption{Juxtaposing Deepcode \& Modulo-SK: BER vs.~feedback $\snr$ for fixed feedforward $\snr=0\dB$. Deepcode uses $K=50$ and Modulo-SK uses $K=13$.
    }
\end{figure*}

\begin{figure*}[!ht]
    \centering
    \begin{subfigure}[t]{0.5\textwidth}
        \centering
        \includegraphics[height=2.7in]{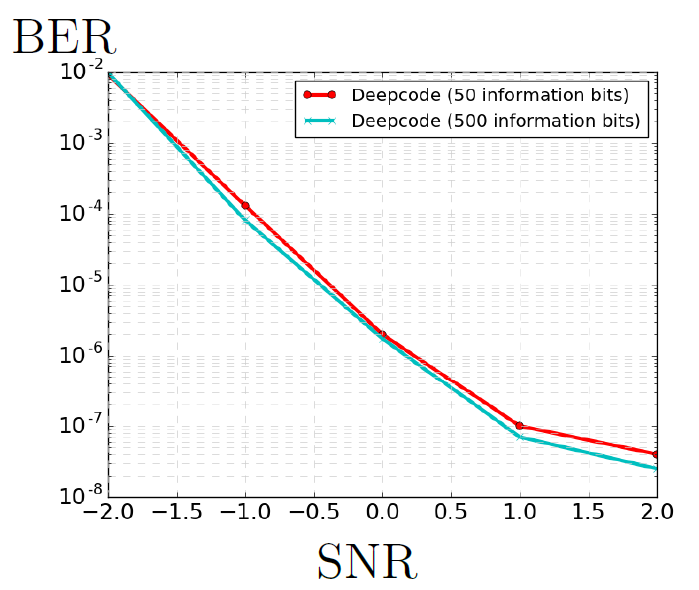}
        \caption{\textbf{Deepcode. \cite[Figure~5 (Middle)]{kim2018deepcode}}
        \label{fig:Figure5middleDeepCode}
        }
    \end{subfigure}%
    ~ 
    \begin{subfigure}[t]{0.5\textwidth}
        \centering
        \includegraphics[height=2.7in]{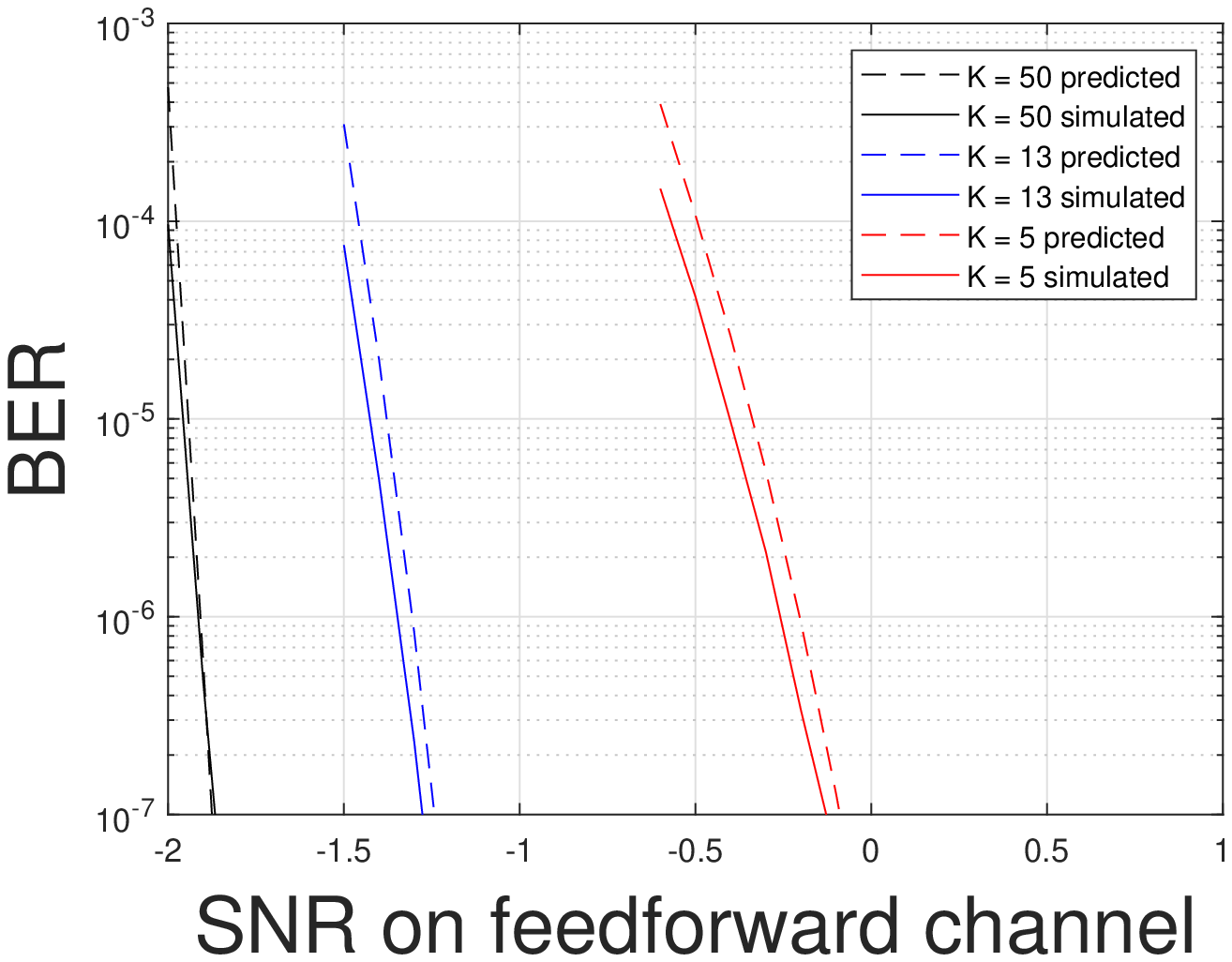}
        \caption{\textbf{Modulo-SK}
        \label{fig:Figure5middleModuloSK}
        }
     \end{subfigure}
    \caption{Juxtaposing Deepcode \& Modulo-SK: BER vs.~feedforward $\snr$. Deepcode uses noiseless feedback, yet Modulo-SK uses noisy feedback with $\snr$ of $27\dB$. \label{fig:Figure5middle}}
\end{figure*}
In this section, we compare the performance of Deepcode to Modulo-SK with no dither. All the comparisons are taken at rate $R=1/3$. 

Figure~\ref{fig:Figure5LeftDeepcode} is copied from \cite[Figure~5 (Left)]{kim2018deepcode}. It depicts the BER 
vs.~feedback $\snr$ attained by Deepcode, for a fixed feedforward $\snr=0\dB$, using $N=150$  iterations (i.e. communication rounds). Figure~\ref{fig:Figure5LeftMSK} depicts BER vs.~feedback $\snr$ for Modulo-SK with the same setting, but using only $N=39$ ($K=13$ information bits). We provide both analytical error upper bounds (which bound the symbol error rate and assume dither) and simulation results (without dither, performing better). Note that by construction (see \cite{ben2017interactive}), Modulo-SK requires high feedback $\snr$ in order to be useful. Nevertheless, once exceeding a certain threshold feedback $\snr$, the slope of the BER curve is much steeper than the one attained by Deepcode. For example, to attain a BER of $10^{-4}$, Deepcode requires a feedback $\snr$ of around $19\dB$, whereas Modulo-SK uses less than $16\dB$, indicating a gap of over $3\dB$. Figure~\ref{fig:Figure5LeftMSK} also shows lower bounds on the BER attainable by any forward error correction code (without feedback) using block lengths of $N=39$ and $N=150$. These bounds were obtained from the finite-block length converse analysis in~\cite{YuryFinite} (calculated using the code provided in \cite{spectre}), further divided by the number of information bits. Note that the BER attained by Deepcode is higher than the corresponding no-feedback lower bound (for $N=150$) over the entire range of feedback $\snr$ levels given in the Deepcode paper (up to $20\dB$). In contrast, note that Modulo-SK attains a BER lower than the corresponding no-feedback lower bound (for $N=39$) with feedback $\snr$ of $15\dB$, thereby proving that noisy feedback can be useful.

Figure~\ref{fig:Figure5middleDeepCode} is copied from \cite[Figure~5 (Middle)]{kim2018deepcode}. It depicts BER vs.~feedforward $\snr$ for Deepcode with noiseless feedback. We compare its performance to Figure~\ref{fig:Figure5middleModuloSK}, which uses the Modulo-SK with \textit{noisy} feedback where the feedback $\snr = 27\dB$. We use three different numbers of rounds: $N = 15$, $N = 39$ and $N = 150$. It can be seen that in this regime, Modulo-SK with noisy feedback significantly outperforms Deepcode, even though the latter uses noiseless feedback. 
For example, at a feedforward $\snr$ of $0\dB$, Deepcode attains an error probability of $10^{-6}$ using $N=150$ with noiseless feedback, whereas Modulo-SK attains a lower error probability using one-tenth of the number of rounds ($N=15$) with noisy feedback. Note that if feedback is noiseless, then Modulo-SK (which in this case essentially coincides with classical SK) can clearly achieve this superior performance even with passive feedback. 
\section{The Issue of Dither}
The classical SK scheme \cite{S-K_partII} assumes a noiseless feedback link and is known to completely fail in the presence of arbitrarily low noise in the feedback link. To remedy that, one can consider exponentially increasing the power over the feedback link to mimic the noiseless case -- but this is of course not practical. Nevertheless, as we showed in \cite{SimpleInteractionAllerton2014,ben2017interactive}, this naive approach can be made practical via the use of modulo-arithmetic, which is a common non-linear technique used to tackle power constraint issues (see for example \cite{Tomlinson},\cite{KochmanZamirJointWZWDP},
\cite{ErezShamaiZamir}).  

In the analysis of the Modulo-SK scheme, we have standardly used a common dither signal known to both terminals, which is added and subtracted on both sides. The purpose of the dither is to ensure that the feedback transmission is uniformly distributed over the modulo cell, which enables calculating its average power. Although it was not emphasized in our papers, it is known (for example, {in high-resolution quantization}) that dither is often not required in practice, and that removing it has a negligible (and sometimes even beneficial) effect on performance. 
Indeed, in the Modulo-SK scheme, the output of the modulo operation is more concentrated than uniform in the first rounds (which is better), and becomes close to uniform in subsequent rounds, even without the dither. The Matlab simulation we provide in \cite{ModuloSKcode} does not use dither, and measures the feedback transmission in order to make sure it does not exceed its power constraint.
\section{Numerical Stability of SK with Noiseless Feedback}
\begin{figure*}[!ht]
    \centering
    \begin{subfigure}[t]{0.55\textwidth}
        \centering
        \includegraphics[height=2.7in]{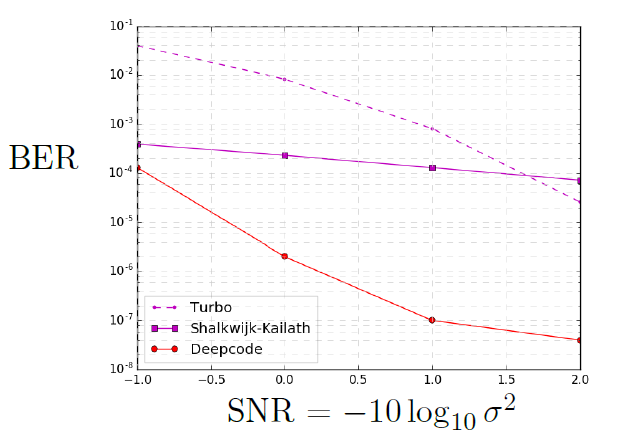}
        \caption{\textbf{From~\cite{{kim2018deepcode}} (\cite[Figure~2 (Left)]{kim2018deepcode}}
        \label{fig:Figure2leftDeepCode}
        }
    \end{subfigure}%
    ~ 
    \begin{subfigure}[t]{0.55\textwidth}
        \centering
        \includegraphics[height=2.7in]{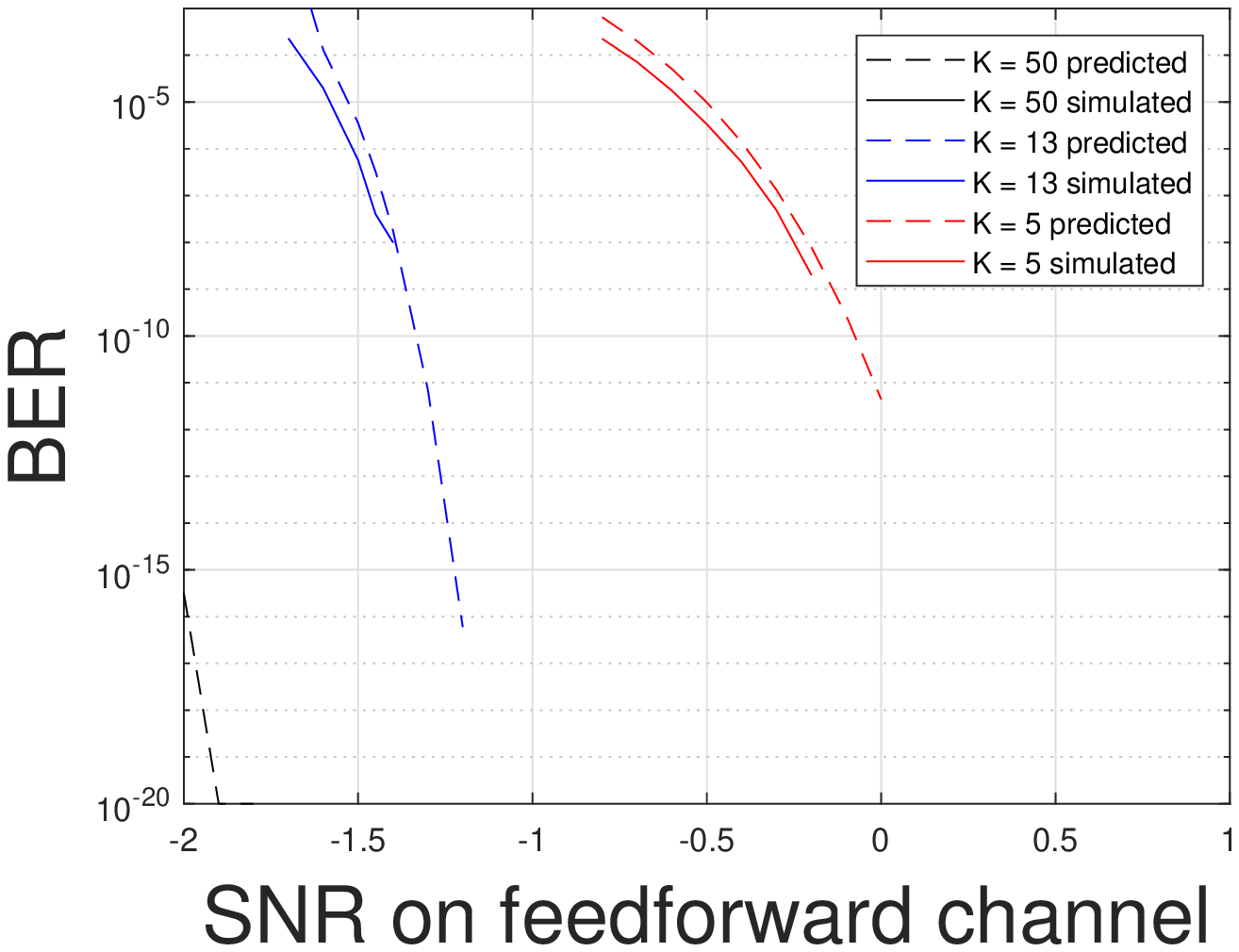}
        \caption{\textbf{Using the code in~\cite{ModuloSKcode}}
        \label{fig:Figure2leftMatlab}
        }
     \end{subfigure}
\caption{Implementations of the SK scheme\label{fig:Figure2left}}
\end{figure*}

As explained in~\cite{ben2017interactive}, the classical SK scheme uses the noiseless feedback in a way that, after $N$ rounds of communication, the effective signal-to-noise ratio of the system is pushed up to $\snr\cdot(1+ \snr)^{N-1}$. At this level of signal-to-noise ratio, one can use very large simple constellations, such as \textit{uncoded} PAM with $2^{NR}$ constellation points, to encode the information bits and facilitate reliable decoding. The idea is that the loss due to using uncoded modulation is fixed, and is divided over the communication rounds, hence the gap to capacity (in $\dB$) is inversely proportional to the block length of the communication, thereby approaching capacity very fast.

In terms of implementation, it should be clear that the receiver must represent its decoding statistic with slightly more than $NR$ bits, simply since it eventually needs to reliably decode a $2^{NR}$-PAM constellation. In other words, the number of bits used by the receiver to store its statistics grows linearly with the number of rounds, much like the case of forward error correction. It was reported in the Deepcode paper that the SK scheme is extremely sensitive to numerical precision issues, and breaks down for example when using $K=50$ information bits, $N=150$ rounds, and a 
floating point Matlab implementation (see Figure~\ref{fig:Figure2leftDeepCode}). We point out that SK can in fact be implemented in a numerically stable way for any number of information bits and iterations, by a judicious fixed point or floating point implementation that takes into account the required resolution of the signals in the system (see \cite[p. 2419]{ben2017interactive}). 

We provide such a numerically stable reference Matlab implementation in~\cite{ModuloSKcode}, which works in Matlab's native double-precision floating point. In Figure~\ref{fig:Figure2leftMatlab}, we depict the performance attained by this implementation of SK, for various numbers of rounds and $\snr$ levels. It can be seen that the BER curves attained are very steep and match the predictions of the theory (see~\cite{ben2017interactive}). Note that due to run-time constraints, the simulations have been limited to error probabilities above $10^{-8}$. 
We also note that due to the fast decay of the error probability observed in Figure~\ref{fig:Figure2leftMatlab}, a large number of rounds is typically not required, hence avoiding the numerical precision issues altogether.


\bibliographystyle{IEEEtran}
{\footnotesize \bibliography{bibreferences}}

\begin{thebibliography}{10}
\providecommand{\url}[1]{#1}
\csname url@samestyle\endcsname
\providecommand{\newblock}{\relax}
\providecommand{\bibinfo}[2]{#2}
\providecommand{\BIBentrySTDinterwordspacing}{\spaceskip=0pt\relax}
\providecommand{\BIBentryALTinterwordstretchfactor}{4}
\providecommand{\BIBentryALTinterwordspacing}{\spaceskip=\fontdimen2\font plus
\BIBentryALTinterwordstretchfactor\fontdimen3\font minus
  \fontdimen4\font\relax}
\providecommand{\BIBforeignlanguage}[2]{{%
\expandafter\ifx\csname l@#1\endcsname\relax
\typeout{** WARNING: IEEEtran.bst: No hyphenation pattern has been}%
\typeout{** loaded for the language `#1'. Using the pattern for}%
\typeout{** the default language instead.}%
\else
\language=\csname l@#1\endcsname
\fi
#2}}
\providecommand{\BIBdecl}{\relax}
\BIBdecl

\bibitem{kim2018deepcode}
\BIBentryALTinterwordspacing
H.~{Kim}, Y.~{Jiang}, S.~{Kannan}, S.~{Oh}, and P.~{Viswanath}, ``Deepcode:
  Feedback codes via deep learning,'' 2018. [Online]. Available:
  \url{https://arxiv.org/abs/1807.00801}
\BIBentrySTDinterwordspacing

\bibitem{Kim2007}
Y.-H. {Kim}, A.~{Lapidoth}, and T.~{Weissman}, ``{The Gaussian Channel with
  Noisy Feedback},'' in \emph{2007 IEEE International Symposium on Information
  Theory}, 2007, pp. 1416--1420.

\bibitem{SimpleInteractionAllerton2014}
{A. Ben{-}Yishai and O. Shayevitz}, ``{The Gaussian Channel with Noisy
  Feedback: Near-Capacity Performance via Simple Interaction},'' in \emph{Proc.
  52nd Allerton Conf. Communication, Control Computing}, Oct. 2014, pp.
  152--159.

\bibitem{DeepcodeIEEE}
H.~{Kim}, Y.~{Jiang}, S.~{Kannan}, S.~{Oh}, and P.~{Viswanath}, ``Deepcode:
  Feedback codes via deep learning,'' \emph{IEEE Journal on Selected Areas in
  Information Theory}, vol.~1, no.~1, pp. 194--206, 2020.

\bibitem{ModuloSKcode}
\BIBentryALTinterwordspacing
{A. Ben{-}Yishai and O. Shayevitz}. {SK and Modulo-SK Matlab Code}. [Online].
  Available: \url{{https://github.com/assafbster/Modulo-SK}}
\BIBentrySTDinterwordspacing

\bibitem{DeepcodeNIPS}
\BIBentryALTinterwordspacing
H.~{Kim}, Y.~{Jiang}, S.~{Kannan}, S.~{Oh}, and P.~{Viswanath}, ``Deepcode:
  Feedback codes via deep learning,'' in \emph{Advances in Neural Information
  Processing Systems 31}, 2018, pp. 9436--9446. [Online]. Available:
  \url{http://papers.nips.cc/paper/8154-deepcode-feedback-codes-via-deep-learning.pdf}
\BIBentrySTDinterwordspacing

\bibitem{ben2017interactive}
{A. Ben{-}Yishai and O. Shayevitz}, ``{Interactive schemes for the AWGN channel
  with noisy feedback},'' \emph{IEEE Transactions on Information Theory},
  vol.~63, no.~4, pp. 2409--2427, 2017.

\bibitem{YuryFinite}
Y.~{Polyanskiy}, H.~V. {Poor}, and S.~{Verdu}, ``Channel coding rate in the
  finite blocklength regime,'' \emph{IEEE Transactions on Information Theory},
  vol.~56, no.~5, pp. 2307--2359, 2010.

\bibitem{spectre}
\BIBentryALTinterwordspacing
A.~{Collins}, G.~{Durisi}, T.~{Erseghe}, V.~{Kostina}, J.~{Ostman},
  Y.~{Polyanskiy}, I.~{Tal}, and W.~{Yang}. Spectre, short packet communication
  toolbox. [Online]. Available: \url{https://github.com/yp-mit/spectre}
\BIBentrySTDinterwordspacing

\bibitem{S-K_partII}
J.~P.~M. Schalkwijk, ``{A coding scheme for additive noise channels with
  feedback part II: Band-limited signals},'' \emph{IEEE Trans. Inf. Theory},
  vol. IT-12, pp. 183--189, Apr 1966.

\bibitem{Tomlinson}
{ M. Tomlinson}, ``{New automatic equalizer employing modulo arithmetic},''
  \emph{Electronics Letters}, vol.~7, no.~5, pp. 138--139, 1971.

\bibitem{KochmanZamirJointWZWDP}
Y.~Kochman and R.~Zamir, ``{Joint Wyner-Ziv/Dirty-Paper Coding by Analog
  Modulo-Lattice Modulation},'' \emph{IEEE Trans. Inf. Theory}, vol.~55, pp.
  4878--4889, 2009.

\bibitem{ErezShamaiZamir}
U.~{Erez}, S.~{Shamai}, and R.~{Zamir}, ``{Capacity and lattice strategies for
  canceling known interference},'' \emph{IEEE Transactions on Information
  Theory}, vol.~51, no.~11, pp. 3820--3833, 2005.

\end{thebibliography}

\end{document}